\documentstyle[mprocl,psfig]{article}

\def\Journal#1#2#3#4{{#1} {\bf #2}, #3 (#4)}

\def\PRL{\em Phys. Rev. Lett.}
\def\PRD{{\em Phys. Rev.} D}

\def\be{\begin{equation}}
\def\ee{\end{equation}}
\def\bea{\begin{eqnarray}}
\def\eea{\end{eqnarray}}
\begin{document}

\title{BH Punctures as Initial Data for General Relativity}

\author{ S. R. Brandt, B. Br\"ugmann }

\address{Max-Plank-Institut f\"ur Gravitationsphysik (Albert-Einstein-Institut)\\
Schlaatzweg 1, 14473 Potsdam, Germany}


\maketitle\abstracts{
We construct initial data for several black holes with arbitrary momenta
and spins by a new method that is based on a compactification of
Brill-Lindquist wormholes. When treated numerically, the method leads to a
significant simplification over the conventional approach which
is based on throats and an isometry condition.
}
  
In this paper we will use the standard formulation of the ADM initial
value problem as given by York\cite{YorkIVP}.

We will use $g_{ab}$ to represent the physical 3-metric, which will be related by
a conformal factor $\psi$ to the flat 3-metric by the relation $g_{ab} = \psi^4 \hat{g}_{ab}$.
Likewise, the physical extrinsic curvature of the spacetime is denoted by $K_{ab}$
and is related to the conformal curvature by $K_{ab} = \psi^{-2} \hat{K}_{ab}$.
Furthermore, we use the relation $K_{a}{}^a = 0$.

\section{Brill and Lindquist Data}
The Brill and Lindquist initial data model is given by
\begin{equation}
\psi = 1+\sum_{(n)}{\frac{m_{(n)}}{2\left|\vec{r}-\vec{r}_{(n)}\right|}},\quad
\hat{K}_{ab} = 0
\end{equation}
There are certain advantages and disadvantages to using this in a numerical
code.\\
\\
{\em Advantages:}\\
1. There is no need for an elliptic solve and
	no possibility of numerical error.\\
2. No need to worry about existence or uniqueness of the solution.\\
{\em Disadvantages:}\\
1. There is a coordinate singularity at $\vec{r}=\vec{r}_{(n)}$.  Numerically,
however, this presents surprisingly little difficulty.  Data with this sort of
singularity present has been evolved previously in 3D codes\cite{Camarda,Bruegmann}.
Furthermore, it is expected that numerical codes of the future will use apparent horizon
boundary conditions to remove regions of the grid inside apparent horizons, and that would
naturally exclude these points.\\
2. The very thing which made this problem so simple, namely time-symmetry $K_{ab}=0$,
severely restricts the set of spacetimes we can construct.  Specifically, we can neither
impart boosts nor spins to the holes.

\section{The Method of Throats and Images}

The strategy for this procedure is to excise spherical regions from the
conformally flat grid, and use an isometry condition to supply the boundary conditions
on the spherical surfaces.

The extrinsic curvature is given by a base form, which has boost and spin parameters.
One adds together a number of these base forms equal to the number of black holes in the
initial value problem to obtain an extrinsic curvature with the appropriate ADM values
on each hole (though these parameters can only truly be identified with spin and boost
in the limit of wide separation).

The base solution to the momentum constraint for a single black hole is:
\begin{equation}
\hat{K}^{ab} = \frac{3}{2 r^2} \left(P^a n^b+P^b n^a-\left(\hat{g}^{ab}-n^a n^b\right) P^c n_c \right)
+ \frac{3}{r^3} \left( e^{acd} S_c n_d n^b+e^{bcd} S_c n_d n^a \right)
\end{equation}
where $r$ is the radius from the black hole's center, $n_a$ is the normalized radius vector for $r$, $P_a$ is the
boost vector for the hole, and $S_a$ is the spin vector.  This base form of the extrinsic curvature does not
yet satisfy the isometry condition at the throats, so we must
now add an infinite number of well-defined ``image terms'' together to accomplish this.
This method has several advantages, and has been
implemented in a variety of settings by Cook et. al\cite{CookEtAl}.
As before, we compare advantages and disadvantages of this method.
\\
{\em Advantages:}\\
1. No coordinate singularities -- the spherical regions which contain them have been excised from the grid.\\
2. $\hat{K}_{ab} \ne 0$, and thus boosts and spins can be given to each black hole separately (in some
sense)\\
{\em Disadvantages:}\\
1. The method is difficult to code, inner boundary conditions present a complication which break most
``off-the-shelf'' elliptic solvers.\\
2. It is more difficult to construct numerical approximations when infinite series are involved. It is
difficult to say anything analytically about these data sets.\\
3. Whether solutions exist and are unique is generally unknown (except for a few cases\cite{Beig}).

\section{A New Method}

We propose a new method\cite{BBivp} which, essentially,
extends the Brill and Lindquist solution by using the base momentum
constraint solution used in the method of throats and images.  The solution is of this form:
\begin{equation}
\psi = u+\frac{1}{\alpha},\quad
\frac{1}{\alpha} = \sum_{(n)} \frac{m_i}{2\left|\vec{r}-\vec{r}_{(n)}\right|},\quad
\beta = \frac{1}{8} \alpha^7 \hat{K}_{ab} \hat{K}^{ab}
\label{neweqn}
\end{equation}
With this ansatz the Hamiltonian constraint can be written as an elliptic
equation for $u$ on all of $R^3$ (no singularities at the
$\vec{r}_{(n)}$), and one can use
a generic elliptic solver to obtain a solution for $u$, which is regular over the entire
grid.\\
{\em Advantages:}\\
1. Can accommodate boosts and spins.\\
2. Simple to code.\\
3. Less difficult to approximate.  We can construct solutions that are first order accurate
in $\beta$ -- the boost/spin term -- for single black holes by expanding
$u$ as $u = 1 + \epsilon u_1+...$ and writing $\epsilon \beta$ instead of just $\beta$ in
Eq.(\ref{neweqn}).
Note that this first order correction is an {\em
upper bound} to the full non-linear solution.
One can easily use this to check whether a given numerical solution makes sense. Any
correct numerical solution must have the property $1 \le u \le u_1$ everywhere on the grid.  In one of the
plots below, a first order approximate analytic solution is compared to a fully non-linear numerically
generated solution for a single boosted and spinning black hole.\\
4. Existence and uniqueness can be proven\cite{Cantor}.  We can also determine that $u$ has
no minima.  Further, if comparing two solutions that differ only in the overall
magnitude of the boost/spin term $\beta$, the solution for $u$ with the larger boost/spin
will be everywhere greater then the solution with the smaller boost/spin.\\
{\em Disadvantages:}\\
1. While there no longer are singularities at the punctures, $u$ is only 
twice differentiable there. With the standard second order stencils,
convergence is not second order near the punctures, which however is
confined to a region close to the punctures. The tiny dip at the punctures
in figs.\ 1 and 2 is a genuine feature, as can be seen for higher
resolution in both the numerics and the approximation.

\begin{tabular}{cc}
\psfig{figure=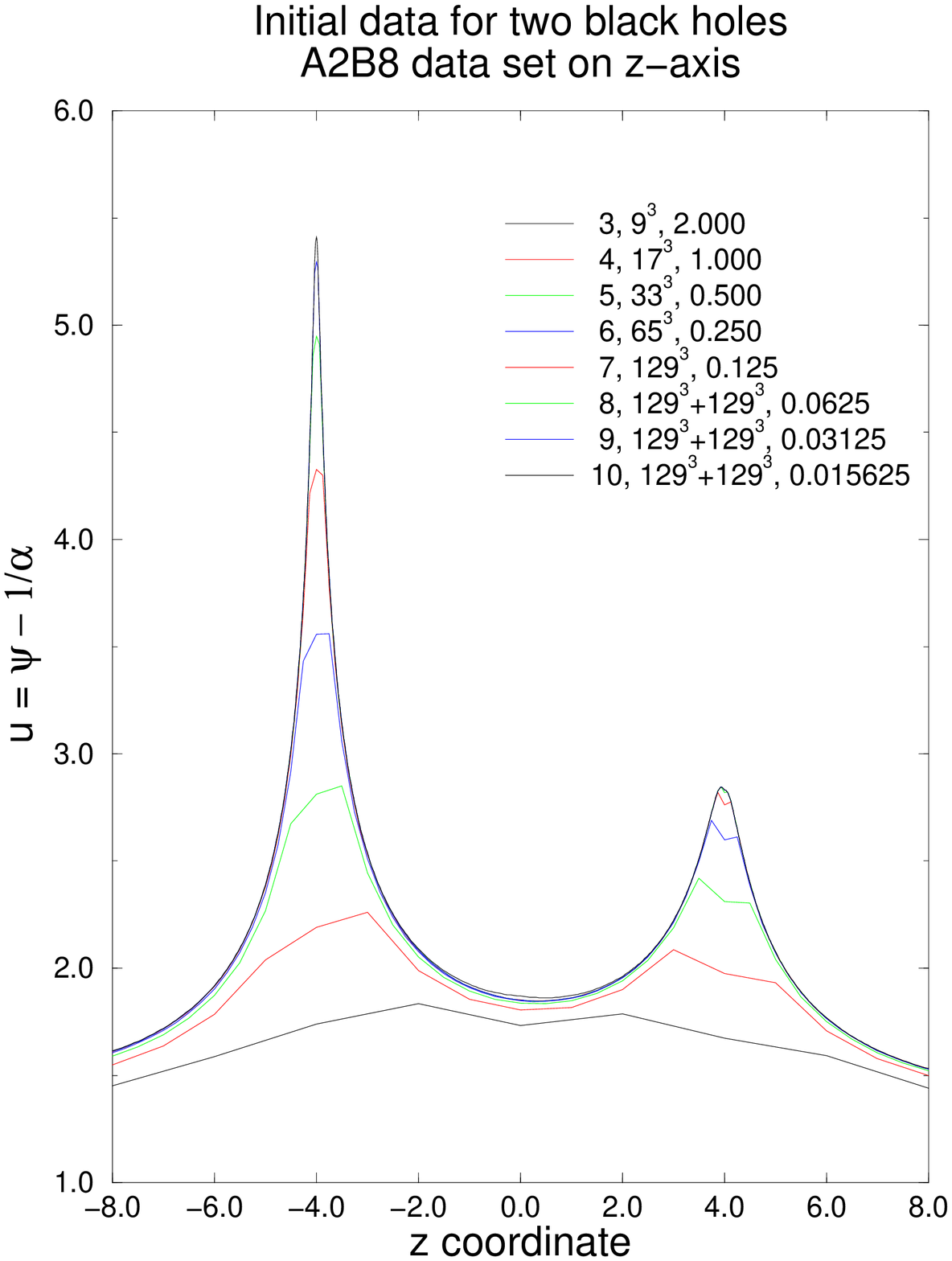,height=5.0cm,width=5.0cm} &
\psfig{file=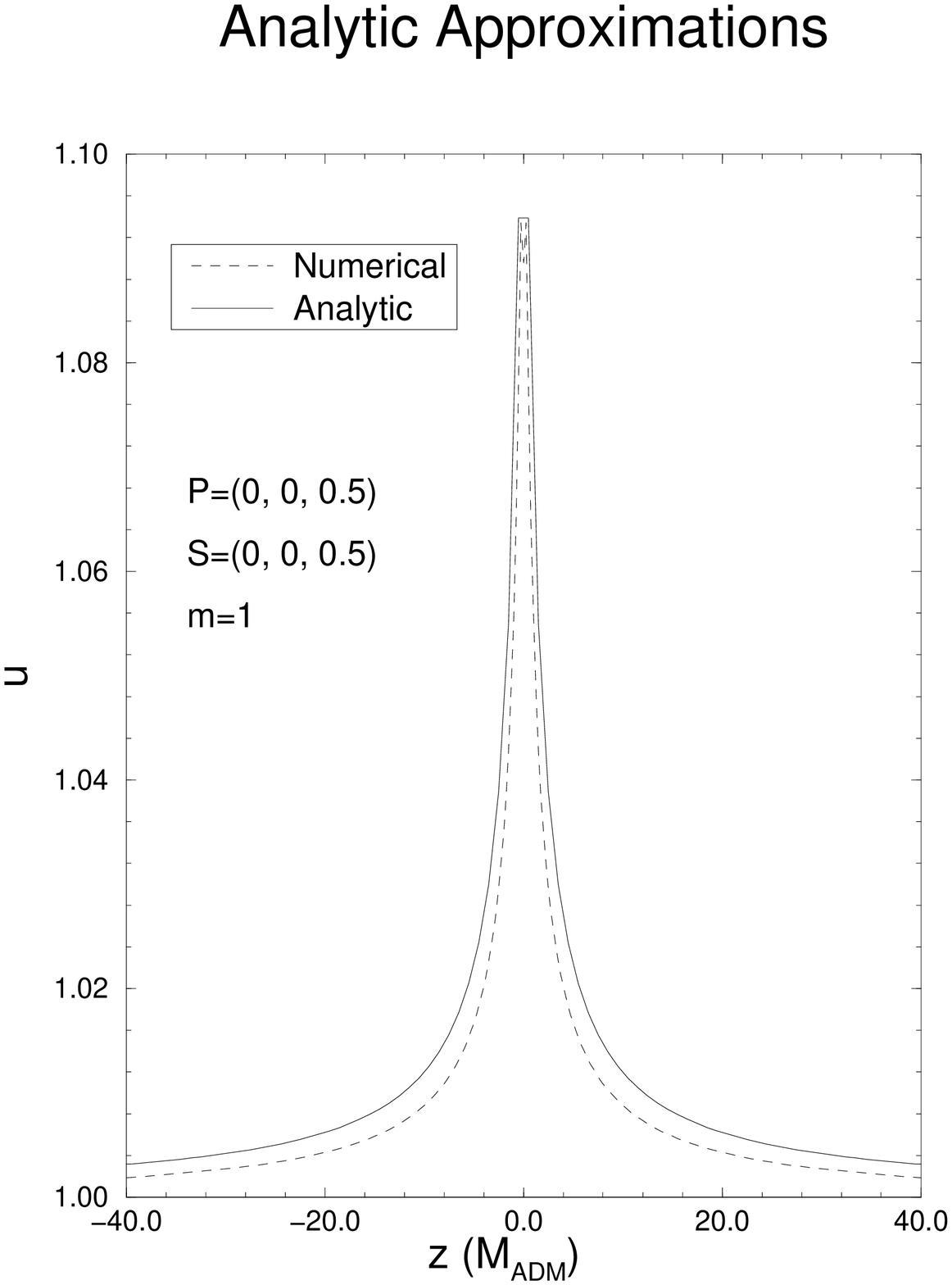,height=5.0cm,width=5.0cm} \\
\begin{minipage}[t]{6 cm}
Figure 1: Convergence study for spinning black holes undergoing spiral infall
\end{minipage}&
\begin{minipage}[t]{6cm}
Figure 2: Comparison between analytic approximation and full numerical solve for
a single boosted spinning black hole
\end{minipage}
\end{tabular}


\section*{References}
\begin{thebibliography}{99}

\bibitem{YorkIVP} James York in {\em Sources of Gravitational Radiation},
ed. L Smarr (Cambridge University Press, Cambridge, England, 1979).

\bibitem{Camarda} P. Anninos, K. Camarda, J. Mass\'o, E. Seidel,
W.-M. Suen and J. Towns
\Journal{\PRD}{52}{2059}{1995}.

\bibitem{Bruegmann}B. Br{\"u}gmann
\Journal{\PRD}{54}{7361}{1996}.

\bibitem{CookEtAl}
G. Cook, M. Choptuik, M. Dubal, S. Klasky, R. Matzner, and
S. Oliveira,
\Journal{\PRD}{47}{1471}{1993}.

\bibitem{BBivp}
S. Brandt and B. Br{\"u}gmann
\Journal{\PRL}{78}{3606}{1997}

\bibitem{Beig}
R. Beig and N. O'Murchadha,
\Journal{\em Class. \& Quant. Grav.}{11}{419}{1994}; \Journal{\em Class. \& Quant. Grav.}{13}{739}{1996}.

\bibitem{Cantor}
M. Cantor,
\Journal{Journal of Mathematical Physics}{20}{1741}{1979}.

\end{thebibliography}
\end{document}